\documentclass[aps,prb,twocolumn,superscriptaddressgroupaddress,floatfix]{revtex4-2}
\usepackage{amssymb}
\usepackage{graphicx}
\usepackage{times}
\usepackage{amsmath}
\usepackage{amsthm}
\usepackage{amsfonts}
\usepackage[T1]{fontenc}
\usepackage[latin9]{inputenc}
\usepackage{array}
\usepackage{multirow}
\usepackage{color}
\usepackage{bm}
\usepackage{color}

\usepackage{bbm}
\usepackage{hyperref}
\usepackage{babel}
\usepackage[mathscr]{euscript}
\usepackage{float}
\usepackage{hyperref}
\hypersetup
{	colorlinks,%
	citecolor=green,%
	linkcolor=red,%
	urlcolor=blue%
}
\allowdisplaybreaks[4]

\begin{document}
\title{Negative intercept of the apparent zero-temperature extrapolated linear-in-\texorpdfstring{$T$}{T} metallic resistivity}
\author{Yi-Ting Tu}
\author{Sankar Das Sarma}
\affiliation{Condensed Matter Theory Center and Joint Quantum Institute, Department of Physics, University of Maryland, College Park, Maryland 20742, USA}

\begin{abstract}
We consider the well-known phonon scattering induced high-temperature linear-in-$T$ metallic resistivity, showing that a naive extrapolation of the effective linearity from high temperatures to $T=0$ leads to an apparent zero-temperature negative resistivity.  The precise magnitude of this extrapolated $T=0$ negative resistivity depends on the temperature regime from where the extrapolation is carried out, and approaches the correct physical result of zero resistivity at $T=0$ only if the extrapolation starts from $T\gg T_D$, where $T_D$ is the Debye temperature.  We establish a theoretical relationship between the negative intercept and the slope of the linear-in-$T$ resistivity as a function of the temperature $T$ from where the extrapolation is carried out.  Experimental implications of our finding are discussed for the much-discussed Planckian behavior of the transport scattering rate.
\end{abstract}

\maketitle

\section{Introduction and Background}\label{sec:intro}

The phonon scattering induced metallic resistivity is among the very first quantum mechanical many-body problems studied in the physics literature, universally referred to as the Bloch-Gr\"uneisen transport theory developed originally almost 100 years ago~\cite{Bloch1930,Gruneisen1933,Ziman2001,Grimvall1999}. A salient feature of this theory is a theoretical explanation for the universally observed linear-in-temperature ($T$) resistivity of essentially all metals at room temperatures as well as an explanation for the rapid fall-off of the temperature-dependent resistivity at low temperatures, all arising from the scattering of (mostly) acoustic phonons by electrons, which is strongly suppressed at low temperatures because of the bosonic phonon occupancy factor.  The basic theory, when combined with accurate electronic band structure details, is in excellent quantitative agreement with the experimental resistivity in many metals~\cite{Allen1978,Pinski1981,Allen1986,Sanborn1989}. The current work deals with aspects of phonon-scattering-induced metallic resistivity, which has not been much discussed, but may have implications for the currently active topic of strange metals and Planckian transport~\cite{Patel2019,Greene2020,Hartnoll2022,Patel2023,Polyak2024,Checkelsky2024}.

It has been known for a long time that the high-temperature phonon-scattering induced metallic resistivity is linear in temperature: $\rho\sim T$ for $T\gg T_D$, where $T_D$ is the typical phonon energy, nominally the Debye temperature, but could also be the Bloch-Gr\"uneisen temperature $T_\mathrm{BG}$ for metals with low electron density.  The phonon temperature scale, $T_\mathrm{BG}=2\hbar k_F$, is defined by the energy of the phonons corresponding to a momentum of $2k_F$, where $k_F$ is the electron Fermi momentum, but in metals $T_D<T_\mathrm{BG}$, with $T_D$ being the ``maximum allowed'' phonon energy and hence the high-temperature scale is set by $T_D$.  We use $T_D$ to denote either the Debye temperature or the Bloch-Gr\"uneisen temperature throughout this paper depending on whichever is lower for the specific system.  Note that with decreasing carrier density, eventually $T_\mathrm{BG}$ becomes the typical phonon scale when $T_\mathrm{BG}<T_D$ is reached as $k_F$ becomes small enough~\cite{Hwang2019,Min2012}.

The specific issue addressed in the current work is how the high-$T$ ($\gg T_D$) linear-in-$T$ metallic resistivity is modified as $T$ decreases from the asymptotic $T\gg T_D$ regime to low temperatures.  The result is, of course, well-established in the lowest temperature (the so-called, Bloch-Gr\"uneisen) regime, $T\ll T_D$, where the phonon-induced metallic resistivity falls off as $T^5$ (or $T^4$) in 3D (or 2D) metals~\cite{Ziman2001,Grimvall1999,Hwang2019}. In this very low-$T$ Bloch-Gr\"uneisen regime, phonon scattering is mostly unimportant as other resistive scattering mechanisms (e.g.\ impurity scattering, which is mostly temperature independent) dominate, and in fact, it is typically a challenge to experimentally observe the predicted Bloch-Gr\"uneisen $T^5$ (3D) or $T^4$ (2D) temperature dependence of the metallic resistivity~\cite{Stormer1990,Hwang2008,Efetov2010}. The interesting regime is, however, the broad intermediate-$T$ regime, $T\sim T_D$ (e.g.\ $T_D/5$ < $T < T_D$), where the phonon-scattering induced metallic resistivity is in the crossover regime, where it does not strictly obey either asymptotic behavior of being $O(T^5)$ for $T\ll T_D$ or $O(T)$ for $T\gg T_D$.  This intermediate crossover regime is the focus of the current work.  We discuss and focus on mostly 3D metals (but present very similar results also for 2D metals for the sake of completeness).  We explicitly consider only acoustic phonon scattering since the metallic resistivity at high-$T$ is almost always dominated by scattering from acoustic phonons.  (Our results and conclusions remain qualitatively valid for optical phonon scattering also with $T_D$ being replaced by the corresponding typical optical phonon energy.)

The interesting theoretical finding we report here is that in a very broad experimentally relevant intermediate-temperature (around $T\sim T_D$) regime, $T_D/5<T<T_D$,  the phonon-induced metallic resistivity actually behaves as $\rho(T)\approx A + BT$, and not simply as $\rho(T)\sim T$, where $A$ and $B$ are weakly temperature dependent, with rather nontrivial behaviors (to be discussed below).  As $T$ becomes large, $A$ vanishes asymptotically for $T\gg T_D$, and $B$ becomes a constant, leading to the well-known $\rho(T) \sim T$ high-$T$ temperature dependence.  But, in a large intermediate temperature regime of experimental relevance $A$ is finite, and, remarkably, negative.  This implies that a low-temperature extrapolation from the high-temperature resistivity would intersect the temperature axis at a finite value of $T$, and would produce an effective (inferred) negative value of $T=0$ resistivity based only on this extrapolation.  We emphasize that the actual (i.e.\ not extrapolated from high $T$) phonon-induced resistivity crosses over eventually (for $T\ll T_D$)  to a power law behavior ($O(T^5)$), and quickly vanishes at $T=0$. Of course, in reality, other contributions to $\rho(T)$ take over at lower temperatures, and typically, $\rho(T)$ saturates to a constant $T=0$ resistivity because of the dominance of resistive impurity scattering effects. 

\begin{figure}
    \centering
    \includegraphics[scale=0.75]{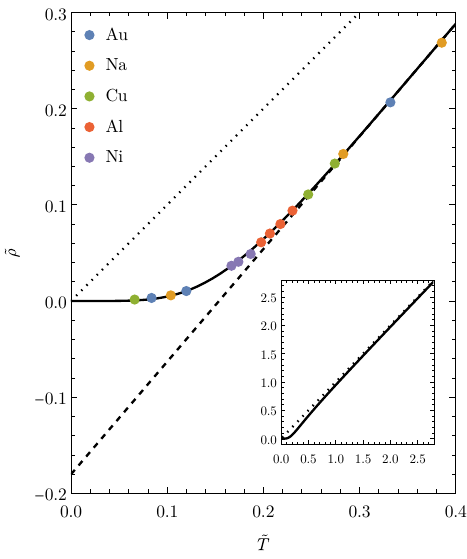}
    \caption{The dimensionless resistivity $\tilde\rho$ as a function of dimensionless temperature $\tilde T$, defined in Eq.~(\ref{eq:dimless}), using the BG formula (solid line). Colored dots show some data for various metals (taken from Fig.~1 of Ref.~\cite{Calatroni2020}). The dashed line indicates the apparent high-$T$ asymptote from the range of the available data, showing a clear negative y-intercept. The dotted line indicates the true asymptote $\tilde\rho=\tilde T$ for $\tilde T\to\infty$. The inset shows the same solid and dotted lines over a larger range of $\tilde T$, showing that $\tilde\rho$ approaches the true asymptote only in a temperature range much larger than the experimental data. }
    \label{fig:measured}
\end{figure}

Fig.~\ref{fig:measured} shows the phonon-induced behavior, $\rho(T) =A + BT$, for a number of regular metals, where the measured resistivity of five metals is shown in dimensionless units, and it is clear that an extrapolation to $T=0$ from the high-$T$ ($T> T_D/5$) resistivity manifests a negative apparent $T=0$ resistivity.  (We also note that the actual, i.e.\ not-extrapolated, resistivity, however, curves up and saturates to a low value, as it should, in contrast to the extrapolated resistivity.)  The problem, however, arises if the resistivity could only be measured at higher values of $T$ (perhaps because the system goes superconducting at some temperature or perhaps because lower $T$ values are not accessible for some reason), then, one could make the perfectly ``reasonable'' extrapolation from the smooth high-$T$ ($>T/T_D>0.2$ in Fig.\ref{fig:measured}) and conclude that the $T=0$ resistivity is negative.  This would be an absurd conclusion.

In reality, one would add the impurity scattering contribution to the resistivity, $\rho_i$, and get $\rho(T) = (\rho_i+A) + BT$, and then an extrapolation to $T=0$ would falsely indicate that the disorder contribution to the $T=0$ resistivity is $\rho_i + A$, which is less than $\rho_i$ since $A$ is negative.  We believe that this ``error'' of estimating the $T=0$ resistivity inadvertently happens often in the experimental literature, particularly when the material goes superconducting at some higher $T$, and an extrapolation is the only way to estimate the effective $T=0$ resistivity. Note that in the fine-tuned, but by no means impossible, scenario of $\rho_i$ and $A$ exactly canceling each other, $\rho_i + A=0$, such an extrapolation would produce the highly misleading conclusion of the system having a perfectly linear-in-$T$ resistivity all the way to $T=0$, incorrectly implying a non-Fermi-liquid ground state. We also note that, although we use acoustic phonon scattering for our explicit calculations, our theoretical considerations apply to all resistive scattering from any bosonic modes in the environment, where such a negative intercept would be generic if the $T=0$ resistivity is inferred from a higher temperature with nominally (but not precisely) linear-in-$T$ resistivity.  This makes our findings relevant to many correlated materials (including cuprates) where a linear-in-$T$ resistivity often manifests with no consensus on the mechanism underlying it, and it is thought that the linearity may arise from scattering by collective bosonic modes associated with a hidden criticality in the system~\cite{Li2024}.

The rest of the work is organized as follows.  In Sec.~\ref{sec:theory}, we provide the theory, establishing the high-$T$ linearity and the low-$T$ Bloch-Gr\"uneisen behavior, providing some analytical expressions for the temperature dependence, both in high-$T$ and low-$T$ limits for both 3D and 2D systems, and providing analytical and numerical details for the putative ``$A + BT$'' behavior of the resistivity in the intermediate temperature regime as mentioned above.  Sec.~\ref{sec:theory} is subdivided into three subsections and presents several graphs providing numerical results compared with the analytical formula in depth.  We conclude in Sec.~\ref{sec:conclusion} with a discussion on the implications of our results for experiments.

\section{Theory and Results}\label{sec:theory}

\begin{figure}
    \centering
    \includegraphics[scale=0.75]{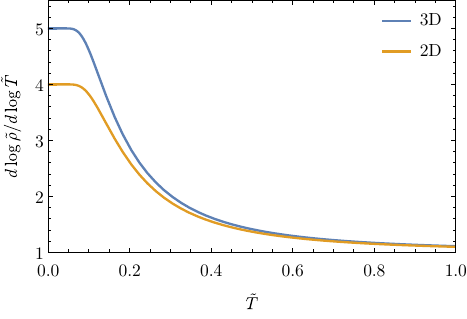}
    \caption{The extracted power of the power of $\tilde T$ in $\tilde\rho$ as a function of $\tilde T$, showing the crossover from the low-$T$ behavior of $T^5$ ($T^4$) in 3D (2D) to the high-$T$ linear-in-$T$ behavior.}
    \label{fig:power}
\end{figure}

We start from the Bloch-Gr\"uneisen (BG) resistivity formula arising from electron-phonon interaction~\cite{Bloch1930,Gruneisen1933,Ziman2001,Grimvall1999,Allen1978,Pinski1981,Allen1986,Sanborn1989,Hwang2019,Min2012}:
\begin{multline}\label{eq:rhoBG}
  \rho(T) = \rho_i \\
  + \frac{2\pi\lambda k_B T/\hbar}{(n/m)e^2}\int_0^{\omega_D} \frac{d\omega}{\omega}{\left(\frac{\omega}{\omega_D}\right)}^4{\left[\frac{\hbar\omega/k_B T}{\sinh(\hbar\omega/2k_B T)}\right]}^2,
\end{multline}
where $n$ is the effective carrier density, and $m$ the effective mass (the exact definitions of $n$ and $m$ are not important in this paper, and our result holds as long as $n/m$ is not strongly $T$-dependent), $\omega_D$ the Debye frequency (corresponding to the Debye temperature $T_D =\hbar \omega_D/k_B$), $\rho_i$ is the ($T$-independent) residual resistivity coming from impurity scattering.

%\subsection{The dimensionless quantities}\label{sec:dimless}

The integral in Eq.~(\ref{eq:rhoBG}) only depends on $T/T_D$ and goes to $1$ when $T/T_D\gg 1$, so it is convenient to rewrite Eq.~(\ref{eq:rhoBG}) as
\begin{equation}\label{eq:dimless}
  \rho(T) = \frac{2\pi\lambda k_B T_D/\hbar}{(n/m)e^2}\cdot \tilde\rho(\tilde T),\quad\tilde T=\frac{T}{T_D},
\end{equation}
where we have the dimensionless version of the BG formula,
\begin{equation}\label{eq:dimlessBG}
  \tilde\rho(\tilde T) = \tilde\rho_i + \tilde T \int_0^{1} \frac{d\tilde\omega}{\tilde\omega}\,{\tilde\omega}^4{\left[\frac{\tilde\omega/\tilde T}{\sinh(\tilde\omega/2\tilde T)}\right]}^2,
\end{equation}
where $\tilde\rho_i$ is the dimensionless $\tilde T$-independent residual resistivity, which we will set to zero in this paper except in Sec.~\ref{sec:rhoi}. We will mainly use the dimensionless variables (indicated by tildes) in this paper.

The BG formula can be generalized to
\begin{equation}\label{eq:general}
  \tilde\rho(\tilde T) = \frac{n-1}{4}\tilde T \int_0^{1} d\tilde\omega\,{\tilde\omega}^{n-2}{\left[\frac{\tilde\omega/\tilde T}{\sinh(\tilde\omega/2\tilde T)}\right]}^2,
\end{equation}
where $n\geq 2$ is an integer. It has the property that $\tilde\rho\sim\tilde T^n$ for $\tilde T\ll 1$ and $\tilde\rho\sim\tilde T$ with unit slope for $\tilde T\gg 1$. The usual 3D case corresponds to $n=5$ and 2D systems have $n=4$. We will focus on these two cases, referring to them simply as 2D and 3D, but note that $n=3$ is also possible for some 3D materials~\cite{Jiang2015}. The crossover of the power of $\tilde T$ in $\tilde\rho$ from the low-$T$ to the high-$T$ behavior is shown in Fig.~\ref{fig:power}.

The integral in Eq.~(\ref{eq:general}) cannot be expressed in terms of elementary function, but one can express it as a finite sum of polylogrithms~\cite{Cvijovic2011}:
\begin{multline}\label{eq:polylog}
\tilde\rho(\tilde T)=(n-1)\,n!\,\zeta(n)\,\tilde T^n\\
-\sum_{j=0}^{n} \frac{(n-1)\,n!}{j!}\,\tilde T^{n-j}\operatorname{Li}_{n-j}(e^{-1/\tilde T}),
\end{multline}
where $\zeta(z)$ is the Riemann zeta function.

\subsection{Asymptotic expansions}

\begin{figure}
    \centering
    \includegraphics[scale=0.75]{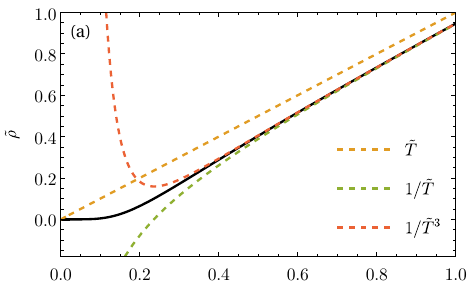}
    \includegraphics[scale=0.75]{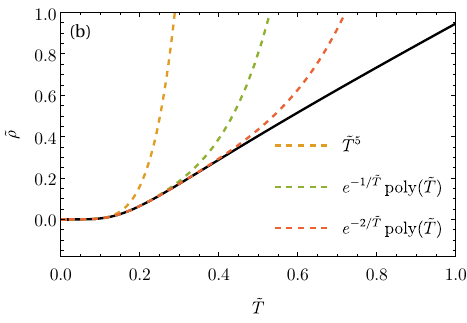}
    \caption{The (a) high- and (b) low-$T$ expansions of $\tilde\rho(\tilde T)$ in 3D up to a given order.}
    \label{fig:expand}
\end{figure}

By expanding the integrand of Eq.~(\ref{eq:general}) in terms of $1/\tilde T$, we have the high-$T$ expansion:
\begin{equation}\label{eq:highT}
    \tilde\rho(\tilde T)=\tilde T - \frac{n-1}{12(n+1)}\,\frac{1}{\tilde T} + \frac{n-1}{240(n+3)}\,\frac{1}{\tilde T^3} + \cdots.
\end{equation}
In particular, in 3D we have
\begin{equation}\label{eq:highT3D}
    \tilde\rho(\tilde T)=\tilde T - \frac{1}{18\tilde T} + \frac{1}{480\tilde T^3} + \cdots,
\end{equation}
and in 2D,
\begin{equation}\label{eq:highT2D}
    \tilde\rho(\tilde T)=\tilde T - \frac{1}{20\tilde T} + \frac{1}{560\tilde T^3} + \cdots.
\end{equation}
The expansions up to the three orders in 3D are plotted in Fig.~\ref{fig:expand}a (the 2D case is visually very similar). The important point is that, to the lowest nontrivial order, we have (take 3D for example---in 2D $4.24$ below is replaced by $4.47$)
\begin{equation}\label{eq:lowest}
    \tilde\rho(\tilde T)\approx \tilde T\left[1-\frac{1}{(4.24\,\tilde T)^2}\right],
\end{equation}
which means that non-linearity only becomes the leading behavior roughly when $\tilde T\lesssim 1/4$. This explains the well-known empirical fact that the metallic linear-in-$T$ resistivity (e.g.\ Fig.~\ref{fig:measured}) persists approximately to $\tilde T\sim 0.25$, and is not limited to $\tilde T\gg 1$ as most metals manifest approximate linear-in-$T$ resistivity down to $50$--$60\,\mathrm{K}$. However, before the second term in Eqs.~(\ref{eq:highT})--(\ref{eq:lowest}) starts to introduce noticeable nonlinearity, the curve is already deviating noticeably from the true asymptote (for $\tilde T\gg 1$) even if it visually looks linear.

Now we turn to the low-$T$ expansion. For $\tilde T \ll 1$, the integrand in Eq.~(\ref{eq:general}) decays exponentially except for $\tilde\omega\ll 1$, and hence we can set the upper limit of the integration to $\infty$, giving~\cite{Cvijovic2011}
\begin{equation}
    \tilde\rho(\tilde T)\approx (n-1)\,n!\,\zeta(n)\,\tilde T^n.
\end{equation}
One way to obtain subleading corrections in elementary functions is to expand the integrand in $e^{-\tilde\omega/\tilde T}$ and integrate over the region that we neglected. That is,
\begin{equation}
    \begin{aligned}
        &\tilde\rho(\tilde T)-(n-1)\,n!\,\zeta(n)\,\tilde T^n\\
        &=-\frac{n-1}{4}\tilde T \int_1^{\infty} d\tilde\omega\,{\tilde\omega}^{n-2}{\left[\frac{\tilde\omega/\tilde T}{\sinh(\tilde\omega/2\tilde T)}\right]}^2\\
        &=-\frac{n-1}{\tilde T}\int_1^{\infty} d\tilde\omega\,\frac{{\tilde\omega}^{n}}{e^{\frac{\tilde\omega}{\tilde T}}}\left(1+2e^{-\frac{\tilde\omega}{\tilde T}}+3e^{-2\frac{\tilde\omega}{\tilde T}}+\cdots\right).
    \end{aligned}
\end{equation}
Now the $k$th term is an elementary integral that gives $e^{-k/\tilde T}$ times a polynomial in $\tilde T$, and therefore we get an expansion in terms of elementary functions. In particular, in 3D we have
\begin{multline}
  \tilde\rho(\tilde T)=480\,\zeta(5)\tilde T^5\\
  -4\,e^{-\frac{1}{\tilde T}}\left(1 + 5 \tilde T + 20 \tilde T^2 + 60 \tilde T^3 + 120 \tilde T^4 + 120 \tilde T^5\right)\\
  -\,e^{-\frac{2}{\tilde T}}\left(4 + 10 \tilde T + 20 \tilde T^2 + 30 \tilde T^3 + 30 \tilde T^4 + 15 \tilde T^5\right)\\
  +\ldots,
\end{multline}
and in 2D,
\begin{multline}
  \tilde\rho(\tilde T)=\frac{4\pi^4}{5}\tilde T^4\\
  -\,e^{-\frac{1}{\tilde T}}\left(3 + 12 \tilde T + 36 \tilde T^2 + 72 \tilde T^3 + 72 \tilde T^4\right)\\
  -\,e^{-\frac{2}{\tilde T}}\left(3 + 6 \tilde T + 9 \tilde T^2 + 9 \tilde T^3 + \frac{9}{2} \tilde T^4\right)\\
  +\ldots.
\end{multline}
The expansions up to the three orders in 3D are plotted in Fig.~\ref{fig:expand}b (again, the 2D case is visually very similar).

\subsection{The apparent asymptote in the intermediate-\texorpdfstring{$T$}{T} regime}

\begin{figure}
    \centering
    \includegraphics[scale=0.75]{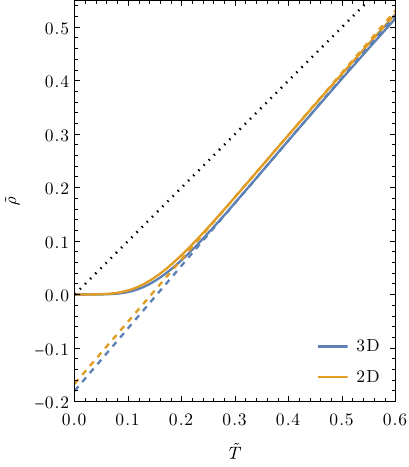}
    \caption{Comparison between dimensionless resistivity $\tilde\rho$ as a function of dimensionless temperature $\tilde T$ for 2D and 3D (solid lines) and their apparent asymptotes (dashed lines). The true asymptote is shown as the dotted black line.}
    \label{fig:rho}
\end{figure}

\begin{figure}
    \centering
    \includegraphics[scale=0.75]{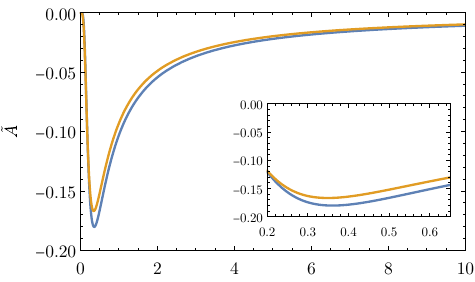}\\
    \hspace{0.1cm}\includegraphics[scale=0.75]{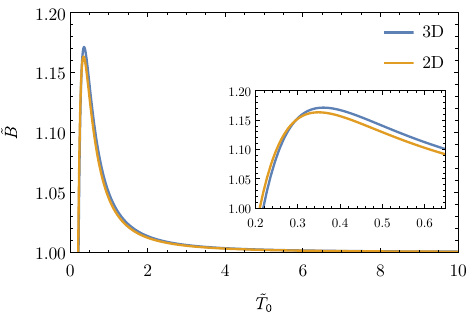}
    \caption{The $\tilde A$ (intercept) and $\tilde B$ (slope) coefficients as a function of the target temperature $\tilde T_0$. Insets show the part of the curves over a smaller and experimentally more relevant regime of $\tilde T_0$.}
    \label{fig:coeffs}
\end{figure}

Note that $\tilde\rho(\tilde T)$ starts to be linear-in-$\tilde T$ around $\tilde T\approx 0.2$ and $\tilde\rho(\tilde T)\sim \tilde T$ (with unit slope) for $\tilde T\gg 1$. However, we can see in Fig.~\ref{fig:measured} that in the experimentally relevant intermediate temperature ($0.2<\tilde T<1$), even if $\tilde\rho(\tilde T)$ already appears to be linear, the tangent line is still very different from the actual asymptote $\tilde\rho=\tilde T$ for $\tilde T\gg 1$.  Here we study the behavior of $\tilde\rho(\tilde T)$ at such intermediate value of $\tilde T$, by approximating
\begin{equation}
  \tilde\rho(\tilde T) \approx \tilde A + \tilde B\tilde T
\end{equation}
around a particular intermediate temperature value $\tilde T \approx \tilde T_0$. For simplicity, we will just use the tangent line. That is, $\tilde A=\tilde\rho(\tilde T_0)-\tilde T_0\tilde\rho'(\tilde T_0)$ is the intercept and $\tilde B=\tilde\rho'(\tilde T_0)$ is the slope of the tangent line, and we will treat $\tilde A$ and $\tilde B$ as functions of $\tilde T_0$ (i.e.\ where the tangent is taken). For actual experimental data, one needs to do a linear fit around a small range of $T$, and hence the best-fit line will not exactly be the tangent line at one particular value of $T_0$. However, since the tangent line only varies slowly with $T_0$ as we will see, this is already enough for our purpose of figuring out how an extrapolation to $T=0$ works when starting from a visually linear higher-temperature ($\sim \tilde T_0$) regime. The corresponding physical coefficients in $\rho(T) \approx A + BT$ are (in 3D)
\begin{equation}
  A = \frac{2\pi\lambda k_B T_D/\hbar}{(n/m)e^2}\tilde A,\quad B = \frac{2\pi\lambda k_B /\hbar}{(n/m)e^2}\tilde B.
\end{equation}

Fig.~\ref{fig:coeffs} shows the numerically calculated $\tilde A$ and $\tilde B$ coefficients as a function of $\tilde T_0$. We can see that, in the regime just above $\tilde T_0\gtrsim 0.2$ where the $\tilde\rho(\tilde T)$ starts to be visually linear, but within the experimentally relevant regime of $\tilde T_0\lesssim 1$, $\tilde A$ and $\tilde B$ only vary slowly with $\tilde T_0$, and have extrema at the inflection point of $\tilde\rho(\tilde T)$. (Note that $\tilde A'(\tilde T_0)=-\tilde T_0\tilde B'(\tilde T_0)$, so their extrema occur at the same point.) The tangent line at that point is special since it visually looks like the asymptote if the curve is only shown up to that range of $\tilde T$ (Fig.~\ref{fig:rho}). And if some set of experimental data is available in that range, attempted linear fits will get very good results (since the curvature vanishes) but the best-fit line will be similar to the local tangent instead of the true asymptote. We call this line the ``apparent asymptote''. For larger $\tilde T_0$, $\tilde A$ slowly approaches $0$ and $\tilde B$ approaches $1$, so that the tangent line approaches the true asymptote $\tilde\rho=\tilde T$ for $\tilde T_0\gg 1$ as it must. However, this is only apparent around $\tilde T_0\sim 5$, which is often beyond the experimentally relevant temperatures. The fact that the asymptotic behavior of the resistivity being strictly linear in $T$ (with no intercept) happens only for $T\gtrsim 5T_D$ although the apparent local linearity commences already near $T \sim T_D/5$ is non-obvious and has not been emphasized in the existing literature.  This implies that any extrapolation to $T=0$ from the apparently linear $T$-dependence in the large intermediate temperature regime of experimental relevance, $T_D/5<T<5T_D$, would lead to an incorrect conclusion about the $T=0$ resistivity.

The numerical values for the apparent asymptote in 3D is
\begin{equation}
    \tilde T_0\approx 0.361,\quad \tilde A\approx 1.171,\quad \tilde B\approx -0.180,
\end{equation}
and in 2D,
\begin{equation}
    \tilde T_0\approx 0.349,\quad \tilde A\approx 1.163,\quad \tilde B\approx -0.167.
\end{equation}
One can obtain an equation of such $\tilde T_0$ from Eq.~(\ref{eq:polylog}). For example, in 3D we have
\begin{multline}\label{eq:exactT0}
  \operatorname{Li}_5(1)
  -\operatorname{Li}_5(x)
  +\log{x}\operatorname{Li}_4(x)
  -\frac{1}{2!}(\log{x})^2\operatorname{Li}_3(x)\\
  +\frac{1}{3!}(\log{x})^3\operatorname{Li}_2(x)
  -\frac{1}{4!}(\log{x})^4\operatorname{Li}_1(x)
  +\frac{1}{5!}(\log{x})^5\operatorname{Li}_0(x)\\
  -\frac{9}{10}\frac{1}{6!}(\log{x})^6\operatorname{Li}_{-1}(x)
  +\frac{21}{10}\frac{1}{7!}(\log{x})^7\operatorname{Li}_{-2}(x)=0,
\end{multline}
where $x=\exp(-1/\tilde T_0)$. The transcendental equation represented by Eq.~(\ref{eq:exactT0}) is not very illuminating, albeit exact, since it cannot be solved analytically. Moreover, this inflection point at $\tilde T_0$ occurs at the intermediate-$\tilde T$ regime where neither the high- nor the low-$T$ expansion can capture its behavior very well, since, by definition, $\tilde T$ is neither large nor small in this important intermediate-$T$ regime. Although the high-$T$ expansion up to $1/\tilde T^3$ does show an inflection point, its position deviates from the exact value by about $30\%$. Therefore, we do not have an accurate perturbative description of the apparent asymptote.

Nevertheless, the apparent negative intercept can be understood intuitively. From Eq.~(\ref{eq:lowest}), linearity only starts around $\tilde T\sim 1/4$. If we estimate the slope slightly above the onset of linearity to be $1$ and extrapolate toward $\tilde T=0$, we should have an intercept that is roughly $-1/4$ (the actual value is closer to $-1/5$). This immediately implies that any extrapolation from the lowest-$T$ regime ($T\sim T_D/5$) where the apparent linearity first manifests, exactly what is typically done experimentally, is bound to provide a large negative intercept for phonon-induced resistivity.  By contrast, an extrapolation from very high-$T$ ($T\gtrsim 5 T_D$) provides the correct result of $\rho (T=0)=0$.  This has serious implications for the residual resistivity at $T=0$, if such a residual resistivity is extracted by extrapolating from a $T>0$ linear-in-$T$ resistivity behavior as discussed below.

\subsection{Extraction of the residual resistivity}\label{sec:rhoi}

\begin{figure}
    \centering
    \includegraphics[scale=0.75]{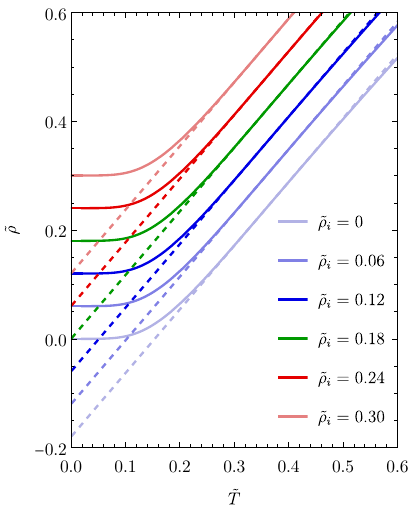}
    \caption{Adding various values of constant $\tilde\rho_i$ to $\tilde\rho(\tilde T)$ in 3D to produce negative (blue), zero (green), or positive (red) values of the apparent intercepts. Dashed lines are the apparent asymptotes.}
    \label{fig:rhoi}
\end{figure}

Here we set the (dimensionless) residual resistivity due to impurity scattering, $\tilde\rho_i$, in Eq.~(\ref{eq:dimlessBG}) to a nonzero value. Note that its relationship to the physical impurity resistivity $\rho_i$ is (in 3D)
\begin{equation}
  \rho_i = \frac{2\pi\lambda k_B T_D/\hbar}{(n/m)e^2}\tilde \rho_i.
\end{equation}
We assume $\rho_i$ to be $T$-independent as it often (but not always~\cite{Ahn2022}) is.

Fig.~\ref{fig:rhoi} shows $\tilde\rho(\tilde T)$ with various values of $\tilde\rho_i$ along with the apparent asymptotes. Since the apparent intercept (i.e.\ the nominal extracted $T=0$ residual resistivity) equals $\tilde\rho_i+\tilde A$, where $\tilde A$ is the negative apparent intercept for zero $\tilde\rho_i$ arising just from phonon scattering, by tuning $\tilde\rho_i$ (which is always positive) we can produce negative, zero, or positive value of the apparent intercept. If experimental data are only available in the range of $0.3\lesssim \tilde T\lesssim 0.6$, one may have misinterpreted the apparent intercept as the residual resistivity due to impurity by linearly extrapolating the data towards zero temperature. Even if data are available up to a larger temperature, one may still be tempted to fit the data linearly (especially if the fluctuation of the data is too large to see the small variation of the slope with $T$), but its intercept will not be a good approximation of $\rho_i$ unless $\tilde\rho_i\gg 0.2$ or $\tilde T\gtrsim 5$. The extrapolated residual resistivity is thus prone to large errors unless the system is so highly disordered that $\rho_i$ is much larger than the magnitude of $A$.

If the effective $T_D$ is known, then one can still get a reasonable approximation of $\rho_i$, by compensating the intercept with the coefficient $A$ at the corresponding temperature range of the available data, even if only a small linear segment of data is available. (We mention that $T_D$ for transport may differ from the $T_D$ inferred from specific heat measurements~\cite{Ziman2001}.)
However, if $T_D$ (or the distribution of $T_D$ and $\lambda$ if multiple phonon modes are effective~\cite{DasSarma2024}) is unknown and when one cannot go to low enough temperature to see the nonlinear-in-$T$ behavior, then it will be difficult to estimate the correct intercept since only $T_0$, not $\tilde T_0$, is known. In this case, if the fluctuation of the data is small enough, one may try to fit the small variation of the slope (that is, the curvature) to extract the phonon parameters, and hence an estimation of $\rho_i$. But if many phonon modes are present, this procedure will only give an effective value of $T_D$ which is roughly, but not always (depending on the relative sizes of $\lambda$ for the individual modes), the largest $T_D$ among the individual modes~\cite{DasSarma2024}. Therefore, an accurate extraction of $\rho_i$ based on only a (nearly-) linear piece of $\rho(T)$ data may not be possible. 
Of course, this problem does not arise if one has transport data down to low enough $T$ values so that the $\rho(T)$ has effectively saturated to $\rho_i$ with the phonon scattering contribution becoming negligible because $T\ll T_D/5$ condition has been achieved.  There are, however, many situations where $\rho(T)$ cannot be measured to arbitrarily low temperatures, and an extrapolation is used to estimate $\rho_i$, which could be problematic because of the effective negative intercept arising from phonon scattering.

\section{Conclusion and Discussion}\label{sec:conclusion}

Using the standard Bloch-Gr\"uneisen transport theory for phonon scattering induced metallic resistivity, we show that the metallic resistivity in a broad range of temperatures ($T_D/5<T<T_D$) manifests a behavior mimicking $\rho=A + BT$, where $A$, $B$ are almost $T$-independent (with $A<0$).  For $T\gg T_D$, $A$ tends to zero (with $B$ becoming strictly $T$-independent) and the resistivity has the well-known strictly proportional to linear-in-$T$ high-$T$ behavior observed in all metals at higher temperatures.  Interestingly, $A$ and $B$, although only weakly $T$-dependent, manifest nontrivial nonmonotonicity with their magnitudes exhibiting extrema as a function of $T$:  both the slope B and the intercept $A$ have a maximum magnitude for $T /T_D\sim 0.36$.  The extremum point (i.e.\ $T/T_D \sim 0.36$) is a nontrivial irrational number arising from the properties of various poly-logarithmic functions in the temperature dependence.  We also show that the high-$T$ linearity has a leading order correction going as $\sim T_D/18T$ and $\sim T_D/20T$ respectively in 2D and 3D, explaining why the apparent approximate linearity in $\rho(T)$ persists to roughly to $T>T_D/5$ since the high-$T$ expansion turns out to be an expansion really in $T/5T_D$ rather than just in $T/T_D$.  But the strict linearity, with $\rho (T)=BT$, can only happen at very high $T/T_D$ where $A$ vanishes and $B$ becomes $T$-independent.

The behavior predicted in this work is routinely observed in the experimental metallic resistivity, as is obvious from our Fig.~\ref{fig:measured} where the resistivity of ordinary metals can be seen to manifest the $A + BT$ type behavior for $T>T_D/5$.  Very similar $T$-dependent resistivity with clear negative intercepts (for $T>T_D/5$) is routinely seen in graphene~\cite{Cao2020,Polshyn2019,Jaoui2022} as well as correlated materials~\cite{Michon2023,Martin1989,Takagi1992}.  We emphasize that the basic behavior of $\rho(T)=A +BT$ in an intermediate temperature regime is not specific to just phonon scattering considered in the current work, but for all resistive scattering from bosonic modes, independent of the type of bosons one considers. The $T_D$ parameter then represents the typical boson energy parameter for the bosons responsible for the resistive scattering. Indeed, such an $A + BT$ behavior in the resistivity (with a negative intercept) can be clearly seen in Fig.~2 of Ref.~\cite{Li2024}, where the resistivity arises from scattering by exotic bosonic modes associated with an underlying quantum critical point.  The only necessity for the manifestation of the negative intercept type behavior is for the system to be pure enough so that the impurity contribution, $\rho_i$,  to the $T=0$ resistivity is smaller than the bosonic scattering contribution at roughly $T \sim T_D/5$ so that $\rho_i$ does not overwhelm the extrapolation of the apparent linear-in-$T$ resistivity.  Of course, the actual (in contrast to the extrapolated) resistivity does not become negative at $T=0$ even for bosonic scattering since at low enough $T$, the resistivity vanishes eventually as a power law.  The problem of the negative intercept is germane only when extrapolating from an apparent linear-in-$T$ resistivity regime $T_D/5<T<T_D$.

The most direct experimental implication of our work is that one should be extremely careful in extracting an effective $T=0$ sample resistivity using extrapolation (as opposed to measurements) from the higher-$T$ regime where the resistivity is roughly linear in $T$, as demonstrated in Fig.~\ref{fig:measured}.  Such an extracted resistivity is producing an effective $T=0$ resistivity given by $\rho_i - |A|$, which is fine when $\rho_i$ is much larger than the magnitude of $A$, i.e., highly disordered samples, but is in fact incorrect quantitatively, and produces an underestimate of the actual $\rho_i$.  This may have implications for the various Planckian behaviors of strange metals much discussed in the literature~\cite{Hartnoll2022,Greene2020}, where the $T=0$ resistivity due to disorder scattering is subtracted out, and often an extrapolation is used from higher-$T$ to estimate the $T=0$ resistivity because of intervening phases (e.g.\ superconductivity) not allowing an actual measurement of the resistivity down to low temperatures.
If an extrapolation of the linear-in-$T$ resistivity is used to estimate the $T=0$ resistivity, then particular care is necessary to ensure that the negative intercept problem discussed in our work is not causing serious errors in the estimated residual resistivity.

We should mention that there is the fine-tuned possibility of the impurity scattering contribution $\rho_i$ to precisely canceling the extrapolated phonon scattering intercept $A$, making $\rho_i +A=0$ in some situations, leading to the incorrect conclusion that the extrapolated $T=0$ resistivity of the system remains linear down to $T=0$.  Since $\rho (T)$ cannot be actually measured down to $T=0$, such extrapolations are often used to infer a non-Fermi liquid behavior of a linear-in-$T$ resistivity down to $T=0$ through extrapolation from a temperature regime where the resistivity is linear---our findings reported in this work suggest that such extrapolated conclusions could be incorrect as a matter of principle.

One important point to note here is that just finding a linear-in-$T$ resistivity over a temperature range, even when it persists to some physically low-temperature scale is not sufficient to claim an underlying non-Fermi liquid behavior which presumes that the scattering rate is linear-in-$T$ even at $T=0$.  One must ask whether the linearity is better fit by a formula of the type $A +BT$, which then immediately implies a possible Fermi liquid independent of whether $A$ is positive or negative.  Subtracting out a hypothetical $A$ (arising from elastic impurity scattering) from the extrapolated resistivity could be misleading because of what we find in this work.  Extrapolations are always dangerous in terms of trying to figure out the functional form in an extrapolated regime outside the measured domain.  A trivial example is the simple analytic function $f(x)= x^2/ (1+x)$, which goes as $x^2$ ($x$) for $x\ll$ ($\gg$)$1$. Any extrapolation from $x\gg1$ is insufficient to figure out the functional form near $x\sim 0$, and the extrapolated value of $f(x)$ for $x=0$ will depend on the regime of $x$ where the extrapolation begins. In this particular (trivial) example, the intercept is $-1$ when extrapolated from the $x\ll 1$ linear regime (where $f(x) \sim x$), but the intercept would be some value between $-1$ and $0$ if the extrapolation is done from a regime where $x\sim 1$ or $<1$.  The problem with phonon scattering is of course nontrivial, but is formally similar to this trivial example. This problem is further compounded by adding a constant to $f(x)$ since now the extrapolation will depend both on this added constant and the origin of the extrapolation.  Only when $f(x)$ extrapolates linearly all the way to zero at $x=0$ can one be sure that $f(x)$ behaves as $x$ near $x\sim0$.  In the presence of an added constant (akin to the impurity scattering contribution), one must take particular care that there is no fine-tuning which leads to such a conclusion.  The real problem is that the resistivity is never measured over a sufficiently large range of linear-in-$T$ regime at low enough temperatures for one to confidently carry out an extrapolation to figure out its effective $T=0$ behavior.
The clear conclusion of our work is that particular caution is necessary in extrapolating a linear-in-$T$ resistivity over some range of $T$ to $T=0$ in forming any conclusion about the actual $T=0$ behavior of resistivity.

\section*{Acknowledgment}
This work is supported by the Laboratory for Physical Sciences through the Condensed Matter Theory Center.

\bibliographystyle{apsrev4-2}
\bibliography{references}

\end{document}